\documentclass{PoS}

\usepackage{mathtools}
\usepackage{tikz-cd}
\usepackage{amsmath}
\usepackage{hhline}
\usepackage[amsmath]{empheq}
\usepackage{cancel}

\usepackage{mathrsfs}
\usepackage{tikz}

\newcommand{\dd}{\mathrm{d}}

\newcommand{\e}{{\epsilon}}
\newcommand{\w}{\wedge}
\newcommand{\bbm}{\left(\begin{matrix}}
\newcommand{\ebm}{\end{matrix}\right)}
\newcommand{\beq}{\begin{eqnarray}}
\newcommand{\eeq}{\end{eqnarray}}
\newcommand{\T}{\text{T}}

\newcommand{\cbral}{[\![}
\newcommand{\cbrar}{]\!]}
\newcommand{\superd}[1]{\,\mathbf d{#1}}

\newcommand{\ali}[1]{\begin{align}#1\end{align}}

\newcommand{\sfrac}[2]{{\textstyle\frac{#1}{#2}}}

\newcommand{\be}{\begin{equation}}
\newcommand{\ee}{\end{equation}}

\newcommand{\beqa}{\begin{eqnarray}}
\newcommand{\eeqa}{\end{eqnarray}} \newcommand{\eq}[1]{(\ref{#1})}
\def\nn{\nonumber} \def \bea{\begin{eqnarray}} \def\eea{\end{eqnarray}}

\newcommand{\barr}{\begin{array}}
\newcommand{\earr}{\end{array}}

 \def\d{\delta} 

    \def\r{\rho}
 \def\S{\Sigma}  

   \def\X{\mathbb X} \def \A{\mathbb A}
 \def \B{\mathbb B}

 \def\one{\mbox{1 \kern-.59em {\rm l}}}

\def\bit{\begin{itemize}} \def\eit{\end{itemize}}

\def\({\left(} \def\){\right)}

\title{BRST symmetry of doubled membrane sigma-models}

\ShortTitle{BRST symmetry of doubled membrane sigma-models}

\author{Athanasios Chatzistavrakidis, Clay J. Grewcoe, \speaker{Larisa Jonke}, Fech Scen Khoo
\\
        Rudjer Bo\v skovi\' c Institute, Zagreb\\
        E-mail: \email{athanasios.chatzistavrakidis@irb.hr}, \email{cgrewcoe@irb.hr}, \email{larisa@irb.hr},  \email{Fech.Scen.Khoo@irb.hr}}

\author{Richard J. Szabo\\
        Heriot-Watt University, Maxwell Institute for Mathematical Sciences, The Higgs Centre for Theoretical Physics, Edinburgh \\
        E-mail: \email{R.J.Szabo@hw.ac.uk}}

\abstract{Courant sigma-models encode the geometric and non-geometric fluxes of compactified closed string theory as generalized Wess-Zumino terms and exhibit their relation to Courant algebroids.  In recent work, we proposed a doubled membrane sigma-model that establishes the corresponding connection to double field theory and its algebroid structure. The strategy is to consider a ``large'' Courant sigma-model over a doubled target spacetime and identify a suitable projection that leads to a sigma-model for doubled fields. In this note, we provide further details for this construction. Starting from the BRST symmetry of the BV action that satisfies the classical master equation, we consistently project the BRST transformations of the superfields of the ``large'' Courant sigma-model to obtain the gauge transformations of the doubled membrane sigma-model. We show that demanding gauge invariance and the closure of gauge transformations of the worldvolume theory, leads to a condition that is in direct correspondence to the strong constraint of the target space double field theory. 

         }

\FullConference{Corfu Summer Institute 2018 "School and Workshops on Elementary Particle Physics and Gravity"\\
		(CORFU2018)\\
		31 August - 28 September, 2018\\
		Corfu, Greece}

\begin{document}

\section{Motivation and introduction}

Double field theory (DFT) \cite{S1, S2, HZ1, HZ2}, seen as an attempt to realize the T-duality of closed string theory  at the level of low-energy supergravity, is based on a  generalized geometry of a tangent bundle extended by 1-forms \cite{Hitchin, Gualtieri}.  This generalized tangent bundle is then equipped with  a  bracket, a symmetric bilinear form and a map to a tangent bundle defining the structure of Courant algebroid \cite{C90, LWX, Sev}. The symmetric bilinear form defines an $O(d,d)$ structure relevant for T-duality on a $d$-dimensional target space, while the symmetries of the generalized tangent bundle unify diffeomorphisms and 2-form gauge transformations of the Kalb-Ramond field. Moreover, the properties of the Courant bracket   are used to  systematically determine  background fluxes of string theory and their Bianchi identities \cite{Halmagyi, Blumenhagen}.

Furthermore,  in Ref. \cite{dee}  Roytenberg used graded geometry to show that given the data of a Courant algebroid one can uniquely construct the Batalin-Vilkovisky (BV) master action for a membrane sigma-model which is a first-order functional for generalized Wess-Zumino terms in three dimensions. (See also Refs. \cite{Park1,ikeda,Park2} for earlier work in the same direction.) This Courant sigma-model belongs to a general class of topological sigma-models of AKSZ type \cite{AKSZ} satisfying the classical master equation.  In this particular case one can show that the conditions for gauge (or more generally BRST) invariance of  the  Courant sigma-model and the on-shell closure of the algebra of gauge transformations  follow from the classical master equation and correspond to the defining axioms of a Courant algebroid.  The  membrane sigma-models were subsequently used for a systematic description of closed strings in non-geometric flux backgrounds  \cite{Halmagyi, Mylonas, ChJL, Watamura,p1}.

In Ref. \cite{p1}, where this contribution is mainly based, we proposed a DFT membrane  sigma-model starting from  a  Courant sigma-model defined  over a doubled target spacetime and adopting a suitable projection. Recall that in Courant algebroids the bundle over a base manifold is extended (``doubled''), while in DFT one doubles the coordinates, i.e. the base space. In order to relate the two approaches, we started from a large Courant algebroid defined over a manifold spanned locally by the  set of doubled coordinates $\{X^i,\widetilde X_i\}$. This naturally introduces an $O(2d,2d)$ structure  indicating that a suitable projection to a subbundle with $O(d,d)$ structure is due. This projection was identified and all Courant algebroid structures were projected accordingly to DFT structures; for instance, the characteristic C-bracket of DFT is obtained in this way from the Courant bracket of the large Courant algebroid. The properties of this bracket were analyzed and used to define a DFT algebroid\footnote{For more details, see the contribution \cite{FSK} to this volume  focusing on the algebroid structure of DFT.}. Moreover, the flux formulation of DFT was used to identify the components of the anchor map and with these data we defined a DFT membrane sigma-model.  Finally, we showed that this  worldvolume theory is gauge invariant only under a certain condition which corresponds to the strong constraint of the target space DFT.  

The gauge transformations of the DFT membrane sigma-model were obtained in \cite{p1} by projecting the standard gauge transformations of the large Courant sigma-model. However, the  latter is the bosonic sector of the classical BV action defined using the BV-BRST formalism after all antifields are set to zero. The master action is defined over a graded manifold  in terms of superfields whose components  include the classical fields, ghosts, ghosts for ghosts and antifields. The classical gauge transformations lift to  the BRST transformations of the superfields, and the BRST invariance of the master action is there by construction---the classical BV action satisfies the classical master equation. 

In Ref. \cite{ZoliDFT}, the classical master action of the large Courant sigma-model was projected to the corresponding DFT action  for projected  superfields. This action does not satisfy  the BV master equation and cannot be constructed using AKSZ theory. This is an expected result, since already at the bosonic level the DFT membrane sigma-model is gauge invariant only up to the worldvolume analogue of the strong constraint, and therefore one cannot expect BRST invariance of the full action.  Here we complete this analysis by explicitly constructing  the BRST transformations for all projected superfield components  of the full DFT membrane sigma-model. 

In Section 2 we review  the gauge symmetries of the Courant sigma-model, both in BV-BRST formalism and at the bosonic level. Then, in section 3 we analyze in detail  the gauge symmetries of the membrane sigma-model for DFT obtained by projecting the BRST symmetry of the large Courant sigma-model.  We show explicitly  how the analogue of the strong constraint appears from the gauge invariance of the equations of motion. In Section 4 we briefly present our conclusions and  outlook.

\section{Gauge and BRST symmetries of the Courant sigma-model}

\subsection{Courant sigma-model as a reducible gauge theory with an open gauge algebra}

First we discuss the gauge symmetries of the Courant sigma-model for a membrane worldvolume $\S_3$, defined over a doubled target space ${\cal M}$. The action functional for the bosonic model is
\be \label{csmif}
S_{\text{C}}[\X,\A,F]=\int_{\S_3} \left( F_I\w\dd \X^I+\sfrac 12 \hat\eta_{\hat I\hat J}\, \A^{\hat I}\w\dd \A^{\hat J}-\rho^I{}_{\hat J}(\X)\,\A^{\hat J}\w F_I+\sfrac 16T_{\hat I\hat J \hat K}(\X)\, \A^{\hat I}\w \A^{\hat J}\w \A^{\hat K}\right)\,,
\ee
where  $I=1,\dots,2d$ is a target space index, $\hat I=1,\dots,4d$ is the bundle index and we have considered scalar fields as components of maps $\X=(\X^I):\S_3\to {\cal M}$, 1-forms $\A\in \Omega^1(\S_3,\X^{\ast}\mathbb{E})$, and an auxiliary 2-form $F\in \Omega^2(\S_3,\X^{\ast}T^{\ast}{\cal M})$, and locally we consider the generalized tangent bundle $\mathbb{E}=\T\cal M\oplus \T^\ast\cal M$.  The fields $(\X^I)=(X^i,\widetilde X_i)$  are identified with the pullbacks of the coordinate functions, $X^i=\X^*(x^i)$ and $\widetilde X_i=\X^*(\tilde x_i)$.
The symmetric bilinear form of the Courant algebroid over $\mathbb{E}$ corresponds to the  $O(2d,2d)$-invariant metric
\be \label{eta}
\hat\eta=(\hat\eta_{\hat I\hat J})=\begin{pmatrix}
	0 & 1_{2d} \\ 
	1_{2d} & 0
 \end{pmatrix}~,
\ee
not to be confused with the $O(d,d)$ metric $\eta$ that will appear later. $\rho^I{}_{\hat{J}}$ are the components of the anchor map $\rho\,:\,\mathbb{E}\to T{\cal M}$ and $T_{\hat{I}\hat{J}\hat{K}}$ are related to a general twist of the Courant algebroid, generating a generalized Wess-Zumino term. For a local basis $(e_{\hat{I}})$ of $\mathbb{E}$, they are identified with $\mathbb{X}^{\ast}\left(\langle e_{\hat{I}},[e_{\hat{J}},e_{\hat{K}}]\rangle\right)$, where $\langle\cdot,\cdot\rangle$ and $[\cdot,\cdot]$ are the non-degenerate symmetric bilinear form and the bracket of the Courant algebroid over $\mathbb{E}$ respectively. 

The action \eqref{csmif} is invariant under
the  following infinitesimal gauge transformations \cite{ikeda} 
\bea\label{gt1}
&&\delta_{(\epsilon\!,\,t)} \X^I=\r^I_{\hat J}\,\epsilon^{\hat J}~,\\[4pt]
\label{gt2}&&\delta_{(\epsilon\!,\,t)} \A^{\hat I}=\dd\epsilon^{\hat I}+\hat\eta^{\hat I\hat N}T_{\hat N\hat J\hat K}\A^{\hat J}\e^{\hat K}-\hat\eta^{\hat I\hat J}\rho^I{}_{\hat J}\,\,t_I~, \\[4pt]
\label{gt3}&&\d_{(\epsilon\!,\,t)}F_I=-\dd t_I-\partial_I\r^J{}_{\hat J}\, \A^{\hat J}\wedge t_J-\epsilon^{\hat J}\partial_I\rho^{J}{}_{\hat J}\,F_J+\sfrac 12\epsilon^{\hat J}\partial_I  T_{\hat I\hat L\hat J}\,\A^{\hat I}\w \A^{\hat L}~,
\eea
where $\e^{\hat I}$ is a scalar gauge parameter, dependent on the worldvolume coordinates, and $t_{I}$ is an additional one-form gauge parameter.{\footnote{Note that these additional gauge invariances were not discussed in Ref. \cite{p1}.}} These transformations define a first-stage reducible gauge symmetry, typical for gauge theories that include differential forms with degree larger than one \cite{Henneaux:1989jq,Gomis:1994he}. 
For completeness, and although this is simpler to do directly in the BV formalism, it is  instructive   to check   the gauge invariance of the field equations of the model and the closure of the algebra of  gauge transformations.
Varying \eqref{csmif} with respect to $F_{I}, \A^{\hat{I}}$ and $\mathbb{X}^{I}$ respectively, we find the field equations
\bea\label{CAeoms}
\label{Feom} && {\cal D}\X^I:= \dd\X^I-\r^I{}_{\hat J}\, \A^{\hat J}=0~,\\[4pt]
\label{Aeom} && {\cal D}\A^{\hat{I}}:=\dd \A^{\hat I}-\hat\eta^{\hat I\hat K}\rho^I{}_{\hat K}F_I+\sfrac 12\hat \eta^{\hat I\hat K}T_{\hat K\hat J\hat L}\A^{\hat J}\w \A^{\hat L}=0~, \\[4pt]
\label{Xeom} && {\cal D}F_{I}:=\dd F_I+\partial_I\r^J{}_{\hat K} \,\A^{\hat K}\w F_J-\sfrac 16 \partial_IT_{\hat J\hat K\hat L}\,\A^{\hat J}\w \A^{\hat K}\w \A^{\hat L}=0~.\eea
Let us now examine how the field equation \eqref{Feom} transforms. We find
\bea
\delta_{(\epsilon\!,\,t)} {\cal D}\X^I=\epsilon^{\hat J}\partial_M\r^I{}_{\hat J}\, {\cal D}\X^M-\hat\eta^{\hat J\hat K}\r^I{}_{\hat J} \r^L{}_{\hat K}\, t_L+\epsilon^{\hat J} \A^{\hat K}(2\r^M{}_{[\hat K}\partial_{\underline M}\rho^I{}_{\hat J]}-\r^I{}_{\hat N}\hat \eta^{\hat N\hat M}T_{\hat M\hat K\hat J})~,\eea
where underlined indices are not antisymmetrized. This directly implies that 
\bea\label{c12}
&& \hat \eta^{\hat J\hat K}\r^I{}_{\hat J} \r^L{}_{\hat K}=0~,\\[4pt] 
\label{c12b}&& 2\r^M{}_{[\hat K}\partial_{\underline M}\rho^I{}_{\hat J]}-\r^I{}_{\hat N}\hat\eta^{\hat N\hat M}T_{\hat M\hat K\hat J}=0\,,\eea
whereupon the field equation transforms covariantly.
Next we examine the transformation of the equation \eqref{Aeom} and obtain
\bea
\delta_{(\epsilon\!,\,t)} {\cal D}\A^{\hat I}
&=&-\hat\eta^{\hat I\hat N}(\partial_MT_{\hat N\hat J\hat K}\epsilon^{\hat K} \A^{\hat J}-\partial_M\r^I{}_{\hat N}\, t_I)\w {\cal D}\X^M+
\hat\eta^{\hat I\hat N}T_{\hat N\hat J\hat K}\epsilon^{\hat K} {\cal D}\A^{\hat J}+{}\nn\\[4pt]
&&+\, \sfrac 12\hat\eta^{\hat I\hat K}\big( 3\r^I{}_{[\hat N}\partial_IT_{\hat J\hat L]\hat K}-\r^I{}_{\hat K}\partial_IT_{\hat N\hat J\hat L}-3T_{\hat K\hat R[\hat N}\hat\eta^{\hat R\hat P}T_{\hat J\hat L]\hat P}\big)\epsilon^{\hat N}\A^{\hat J}\w \A^{\hat L}~,\eea
where we used the condition in \eqref{c12b}. We observe that the field equation transforms covariantly provided one more condition holds, namely
\bea\label{c3}
3\r^I{}_{[\hat N}\partial_IT_{\hat J\hat L]\hat K}-\r^I{}_{\hat K}\partial_IT_{\hat N\hat J\hat L}-3T_{\hat K\hat R[\hat N}\hat\eta^{\hat R\hat P}T_{\hat J\hat L]\hat P}=0~.\eea
 It is then easily confirmed that transforming the field equation \eqref{Xeom} does not produce any further conditions. Moreover, the three conditions \eqref{c12}, \eqref{c12b} and \eqref{c3} are precisely the local coordinate expressions for the three independent axioms of a Courant algebroid. 

Closure of the algebra of gauge transformations gives
\bea
[\delta_{(\epsilon_1,\,t_1)},\delta_{(\epsilon_2,\,t_2)}]\X^I&=&\r^I{}_{\hat J}\, \,\epsilon_{12}^{\hat J}~,\\[4pt] \epsilon_{12}^{\hat I}&:=&\hat\eta^{\hat I\hat J}T_{\hat J\hat K\hat L}\epsilon_1^{\hat K}\epsilon_2^{\hat L}~,\eea
where we used the condition in \eqref{c12b} to define $\epsilon_{12}$. 
 Furthermore we have
 \bea\label{ccA}
[\delta_{(\epsilon_1,\,t_1)},\delta_{(\epsilon_2,\,t_2)}]\A^{\hat I}&=&\delta_{(\epsilon_{12},\,t_{12})} \A^{\hat I}-\hat\eta^{\hat I\hat J}\partial_MT_{\hat J\hat K\hat L}\epsilon_1^{\hat K}\epsilon_2^{\hat L} {\cal D}\X^M~,\\[4pt]
t_{12I}&:=&\partial_IT_{\hat K\hat L\hat J}\,\epsilon_1^{\hat K}\epsilon_2^{\hat L}\A^{\hat J}+2\partial_I\r^J{}_{\hat K}\,\epsilon^{\hat K}_{[1}\,\,t_{2]J}~,\eea
where we used the conditions in \eqref{c12b} and \eqref{c3}. The closure on the field $F_{I}$ does not introduce any further conditions.
Therefore we conclude that the  Courant sigma-model is gauge invariant {\it on-shell}, provided that Eqs. (\ref{c12},\,\ref{c12b},\,\ref{c3}) hold. (Sometimes this is referred to as a reducible gauge theory with an open gauge algebra.)

\subsection{The BV action and BRST transformations}

On-shell closure of the algebra of gauge transformations implies that the natural description of the gauge symmetries for the Courant sigma-model is the BV-BRST formalism (for physics-oriented reviews, see \cite{Henneaux:1989jq,Gomis:1994he}).  In particular, one can construct the classical master action \cite{dee}
 \bea \label{master}
{\bf S}_{\text{C}}[{\bf X},{\bf A},{\bf F}]=\int_{T[1]\S_3}  \! \! \mu\left( {\bf F}_I\,\superd{\bf X}^I+\sfrac 12 \hat\eta_{\hat I\hat J}\,{\bf A}^{\hat I}\superd{\bf A}^{\hat J}-\rho^I{}_{\hat J}({\bf X}){\bf A}^{\hat J}\,{\bf F}_I+\sfrac 16T_{\hat I\hat J\hat K}({\bf X}){\bf A}^{\hat I}{\bf A}^{\hat J}{\bf A}^{\hat K}\right)~,
\eea
where $\mu\equiv d^3\sigma d^3\theta$ is the Berezinian measure on the graded manifold  $T[1]\Sigma_3$ spanned by  coordinates $(\sigma^\mu,\theta^\mu)$ of degrees ($0,1$) respectively,   $\superd{}=\theta^\mu\partial_\mu$ is the superworldvolume differential    and  superfields include the classical fields ($\X, \A, F$), ghosts ($\epsilon,t,v$) of ghost numbers ($1,1,2$) and antifields:
\bea\label{sf}
{\bf X}^I &=& \X^I+F^{\dagger I}+t^{\dagger I}+v^{\dagger I}~,\\
{\bf A}^{\hat I}&=&\epsilon^{\hat I}+\A^{\hat I}+\hat\eta^{\hat I\hat J}\A^\dagger_{\hat J}+\hat\eta^{\hat I\hat J}\epsilon^\dagger_{\hat J}~,\\
{\bf F}_I&=&v_I+t_I+F_I+\X^{\dagger}_I~.\eea
Here ${\bf X}^I, {\bf A}^{\hat I}, {\bf F}_I$ are superfields with total degree $0,1,2$ respectively, where the total degree of a field ${\phi}$ is the sum of its ghost number gh($\phi$) and its form degree deg($\phi$). Antifields are denoted by a dagger $\dagger$ and we have gh($\phi$)\,+\,gh($\phi^{\dagger})=-1$ and deg($\phi$)\,+\,deg($\phi^{\dagger})=3$.

The conditions given in Eqs. \eqref{c12}, \eqref{c12b} and \eqref{c3} are obtained directly from the classical master equation $\{{\bf S}_{\text{C}},{\bf S}_{\text{C}}\}=0$, where the (anti)bracket arises from the target manifold symplectic structure
\be\label{sp}
\omega=\dd { \X}^I\,\dd {F}_I+\sfrac 12\hat \eta_{\hat I\hat J}\,\dd{\A}^{\hat I}\dd{\A}^{\hat J}~.\ee
Setting all ghosts and antifields to zero in the master action \eqref{master} reproduces the Courant sigma-model \eqref{csmif}, while the BRST transformations of the classical fields give the gauge transformations as in \eqref{gt1}--\eqref{gt3}.  For completeness we present here the BRST transformations of all the fields,
\ali{
\delta\X^I&=\rho^I_{\hat{I}}\epsilon^{\hat{I}}~,\\
\delta \A^{\hat{I}}&=\dd{\epsilon}^{\hat{I}}-\hat \eta^{\hat{I}\hat{J}}\rho^I_{\hat{J}}t_I+\hat\eta^{\hat{I}\hat{J}}T_{\hat{J}\hat{K}\hat{L}} \A^{\hat{K}}\epsilon^{\hat{L}}-\hat\eta^{\hat{I}\hat{J}}\partial_J\rho^I_{\hat{J}} F^{\dagger J}v_I+\sfrac 1 2 \hat\eta^{\hat{I}\hat{J}}\partial_JT_{\hat{J}\hat{K}\hat{L}} F^{\dagger J}\epsilon^{\hat{K}}\epsilon^{\hat{L}}~,\\
\delta F_I&=-\dd{t}_I-\partial_I\rho^J_{\hat{I}}\epsilon^{\hat I} F_J-\partial_I\rho^J_{\hat{I}} \A^{\hat I}t_J+\sfrac 1 2 \partial_IT_{\hat{I}\hat{J}\hat{K}}\epsilon^{\hat{I}} \A^{\hat{J}} \A^{\hat{K}}+{}\nn\\
&\phantom{\,=\,}+\sfrac 1 2 \partial_IT_{\hat{I}\hat{J}\hat{K}}\hat\eta^{\hat{K}\hat{L}}\epsilon^{\hat{I}}\epsilon^{\hat{J}} \A^\dagger_{\hat{L}}-\partial_I\rho^J_{\hat{I}}\hat\eta^{\hat{I}\hat{J}} \A^{\dagger}_{\hat{J}}v_J+\sfrac 1 2 \partial_I\partial_J\partial_K\rho^L_{\hat{I}} F^{\dagger J} F^{\dagger K}\epsilon^{\hat{I}}v_L-\partial_I\partial_J\rho^K_{\hat{I}}t^{\dagger J}\epsilon^{\hat{I}}v_K-{}\nn\\
&\phantom{\,=\,}-\partial_I\partial_J\rho^K_{\hat{I}} F^{\dagger J}\epsilon^{\hat{I}}t_K
+\partial_I\partial_J\rho^K_{\hat{I}} F^{\dagger J} \A^{\hat{I}}v_K-\sfrac 1 {12} \partial_I\partial_J\partial_KT_{\hat{I}\hat{J}\hat{K}} F^{\dagger J} F^{\dagger K}\epsilon^{\hat{I}}\epsilon^{\hat{J}}\epsilon^{\hat{K}}+{}\nn\\
&\phantom{\,=\,}+\sfrac 1 6 \partial_I\partial_JT_{\hat{I}\hat{J}\hat{K}}t^{\dagger J}\epsilon^{\hat{I}}\epsilon^{\hat{J}}\epsilon^{\hat{K}}-\sfrac 1 2 \partial_I\partial_JT_{\hat{I}\hat{J}\hat{K}} F^{\dagger J} \A^{\hat{I}}\epsilon^{\hat{J}}\epsilon^{\hat{K}}~,\\
\delta\epsilon^{\hat{I}}&=\hat\eta^{\hat{I}\hat{J}}\rho^I{}_{\hat{J}}\,v_I-\sfrac 1 2 \hat\eta^{\hat{I}\hat{J}}T_{\hat{J}\hat{K}\hat{L}}\epsilon^{\hat{K}}\epsilon^{\hat{L}}~,\\[4pt]
\delta t_I&=\dd{v}_I-\partial_I\rho^J{}_{\hat{I}}\epsilon^{\hat{I}}t_J+\partial_I\rho^J{}_{\hat{I}} \,\A^{\hat{I}}v_J-\sfrac 1 2 \partial_IT_{\hat{I}\hat{J}\hat{K}}\epsilon^{\hat{I}}\epsilon^{\hat{J}}\,\A^{\hat{K}}+
\partial_I\partial_J\rho^K{}_{\hat{I}}\,F^{\dagger J}\epsilon^{\hat{I}}v_K -\sfrac 1 6 \partial_I\partial_JT_{\hat{I}\hat{J}\hat{K}}\,F^{\dagger J}\epsilon^{\hat{I}}\epsilon^{\hat{J}}\epsilon^{\hat{K}}~,\label{vart}\\
\delta v_I&=-\partial_I\rho^J{}_{\hat{I}}\epsilon^{\hat{I}}v_J+\sfrac 1 6 \partial_IT_{\hat{I}\hat{J}\hat{K}}\epsilon^{\hat{I}}\epsilon^{\hat{J}}\epsilon^{\hat{K}}\label{varv}~.
}
Note that one needs to introduce a  ghost for ghost  $v$ because we are dealing with a first-stage reducible gauge theory, or said differently, there are ``gauge invariances''  for  gauge transformations  typical for gauge theories including higher differential forms.

\section{Gauge symmetries of the DFT membrane sigma-model}

\subsection{DFT membrane sigma-model}

In Ref. \cite{p1} we showed that one can define a DFT algebroid structure and a corresponding  membrane sigma-model starting from a large Courant algebroid over a $2d$ dimensional space ${\cal M}$ with local coordinates $\{X^i,\widetilde X_i\}$ and applying a suitable projection. 
In particular, we considered  sections $\A$ of the large Courant algebroid $\mathbb{E}$, 
decomposed in a suitable basis,
\bea\label{st}
 \A&=& \A_+^Ie_I^++ \A_-^Ie_I^-~, \\[4pt] e^{\pm}_I&=&\partial_I\pm \eta_{IJ}\,\dd\X^J~,
\\[4pt]  \A_{\pm}^I  &=&\sfrac 12(  \A^I\pm\eta^{IJ}\,\widetilde  \A_J)~,\eea
and projected to the subbundle $L_+$ spanned by the local sections $(e^+_I)$.
Projection of the symmetric bilinear form of $\mathbb{E}$, leads to the $O(d,d)$ invariant DFT metric:{\footnote{Denoting $\A_+=A$ and $\B_+=B$.}}
\be
\langle  \A, \B\rangle_{\mathbb{E}}=\sfrac 12 \hat\eta_{\hat{I}\hat{J}} \A^{\hat{I}} \B^{\hat{J}}=\eta_{IJ}( \A^I_+ \B^J_+- \A^I_- \B^J_-)\quad \mapsto \quad \eta_{IJ}A^IB^J=\langle A,B\rangle_{L_+}~.\ee
This works for general Courant algebroids over ${\cal M}$ with anchor $\rho^I{}_{\hat{J}}=(\rho^{I}{}_J,\widetilde{\rho}^{IJ})$, yielding a  C-bracket: 
\be \label{cbragen}
\cbral A,B\cbrar^J= (\rho_+)^L{}_I\left(A^I\partial_L B^J-\sfrac 12 \eta^{IJ}A^K\partial_LB_K - (A\leftrightarrow B)\right)+\hat{T}_{IK}{}^{J}A^IB^K~,
\ee
in terms of a map $\rho_+:L_+\to T{\cal M}$ with components $(\rho_{\pm})^I{}_J=\rho^I{}_J\pm \eta_{JK}\widetilde{\rho}^{IK}$, and $\hat T$ chosen as:
\be
\hat T_{IJK}:=\sfrac 1 2 T_{IJK}=\sfrac 1 2 \left(A_{IJK}+3B_{[IJ}{}^L\eta_{K]L}+3C_{[I}{}^{LM}\eta_{J\underline{L}}\eta_{K]M}+D^{LMN}\eta_{I[\underline{L}}\eta_{J[\underline{M}}\eta_{K]N}\right)~,
\ee
where $A$, $B$, $C$ and $D$ are the components of $T_{\hat{I}\hat{J}\hat{K}}$:{\footnote{These are not precisely identified with the fluxes of DFT, thus we do not use the corresponding notation $(H,f,Q,R)$. The flux identification is explained in detail in Ref. \cite{p1}.}}
\be
T_{\hat{I}\hat{J}\hat{K}}:=\begin{pmatrix}A_{IJK}&B_{IJ}{}^K\\
	C_{I}{}^{JK}&D^{IJK} \end{pmatrix} \ .\label{Tmatrix}
\ee

Using these projected data we proposed the following DFT membrane sigma-model:
\bea \label{dft1}
S_{\text{DFT}}[\X,A,F]=\int_{\S_3}  \left( F_I\w\dd\X^I+\eta_{IJ}A^I\w\dd A^J-(\rho_+)^I{}_{J}A^J\w F_I+\sfrac 13\hat T_{IJK}A^I\w A^J\w A^K\right)~.
\eea
Next, in parallel to the flux formulation of DFT \cite{S2, dftflux2, dftflux1, dftflux3, dftflux4, dftflux5} we took a parametrization of the $\rho_+$ components to be 
\be\label{dftrho}
(\rho_+)^I{}_J=\begin{pmatrix} \d^i{}_j & \beta^{ij} \\ B_{ij} &\d_i{}^j+\beta^{jk}B_{ki} 
	\end{pmatrix}~.
\ee
 In particular this means that 
\be\label{cd1}
\eta^{JK}(\rho_+)^I{}_J(\rho_+)^L{}_K=\eta^{IL}~,\ee
which is to be compared with the condition in \eqref{c12}.  Moreover, in Ref. \cite{p1} we  proposed a set of infinitesimal gauge transformations 
\bea\label{gtold}
&&\delta_\epsilon \X^I=\r^I{}_J\epsilon^J~,\\[4pt]
&&\delta_\epsilon A^I=\dd\epsilon^I+\eta^{IN}\hat T_{NJK}A^J\e^K~, \\[4pt]
&&\d_{\epsilon}F_I=-\epsilon^J\partial_I\rho^{K}{}_{J}F_K+\epsilon^JA^K\w A^L\partial_I \hat T_{KLJ}~,
\eea
and showed that the action \eqref{dft1} is invariant under these transformations   provided that 
\bea\label{cd3a}
&&2\rho^{K}{}_{[L}\partial_{\underline{K}}\rho^{I}{}_{M]}-\rho_{K[L}\partial^I\rho^K{}_{M]}=\rho^I{}_J\eta^{JK}\hat T_{KLM}~,\\[4pt]
&&3\r^J{}_{[K}\partial_{\underline J}\hat T_{MM']N}-\r^J{}_N\partial_J\hat T_{KMM'}-3\eta^{PJ}\hat T_{P[MM'}\hat T_{K]NJ}=0~.\label{cd3b}\eea
However, these conditions are not sufficient; one needs to additionally impose the following constraint: 
\be\label{anomaly} 
\rho_{KL}\,\partial^I\rho^K{}_M\,\epsilon^M\,F_I
=0~. 
\ee   
As argued in Ref. \cite{p1} this is the way that the strong constraint of the target space DFT appears in the worldvolume theory. 

\subsection{Projecting superfields}

The bosonic action \eqref{dft1} is lifted to the full action in terms of superfields \cite{ZoliDFT}:{\footnote{Writing $\pm$ subscripts explicitly again.}}
\be \label{masterDFT}
{\bf S}_{\text{DFT}}[{\bf X},{\bf A}_+,{\bf F}]=\int_{T[1]\S_3} \mu\left( {\bf F}_I\,\dd{\bf X}^I+ \eta_{IJ}{\bf A}_+^I\dd{\bf A}_+^J-(\rho_+)^J{}_{I}({\bf X}){\bf A}_+^I{\bf F}_J+\sfrac 13\hat T_{IJK}({\bf X}){\bf A}_+^I{\bf A}_+^J{\bf A}_+^K\right)~,
\ee
where in comparison with \eqref{master} we used the structures $(\rho_+,\hat T, \eta)$  of a DFT algebroid and projected the superfield ${\bf A}\to{\bf A_+}$:
\bea\label{sfDFT}
{\bf A}_+^I&=&\epsilon_+^I+\A_+^I+\eta^{IJ}\A^\dagger_{+J}+\eta^{IJ}\epsilon^\dagger_{+J}~,\eea
by setting the $\A_-$  and $\epsilon_-$ to zero, an operation whose consistency will be addressed below.
Next, we project the BRST transformations of the superfields of the large Courant sigma-model \eq{master}. By splitting and projecting the BRST transformation of the field $ \A^{\hat I}$, one obtains:
\ali{
\delta \A^I_+&=\dd\epsilon_+^I-\sfrac 1 2 \eta^{IJ}\rho^K_{+J}\, t_K+ \eta^{IL}\hat T_{LJK}\A^J_+\epsilon^K_+-\sfrac 1 2 \eta^{IJ}\partial_K\rho_{+J}^LF^{\dagger K}v_L+\sfrac 1 2 \eta^{IJ}\partial_K\hat T_{JLM}F^{\dagger K}\epsilon_+^L\epsilon_+^M~,\label{deltaa}\\[4pt]
\delta \A^I_-&=\sfrac 1 2 \eta^{IJ}\rho^K_{-J}\, t_K+\sfrac 1 2 \eta^{IL}\theta_{JKL}\,\A^J_+\epsilon^K_++\sfrac 1 2 \eta^{IJ}\partial_K\rho_{-J}^LF^{\dagger K}v_L+\sfrac 1 4 \eta^{IJ}\partial_K\theta_{LMJ}F^{\dagger K}\epsilon_+^L\epsilon_+^M~,
}
and for the ghost field $\epsilon^{\hat I}$:
\ali{
\delta \epsilon^I_+&=\sfrac 1 2 \eta^{IJ}\rho_{+J}^Kv_K-\sfrac 1 2 \eta^{IL}\hat T_{LJK}\epsilon^J_+\epsilon^K_+~,\label{deltaepsilon}\\[4pt]
\delta \epsilon^I_-&=-\sfrac 1 2 \eta^{IJ}\rho_{-J}^Kv_K-\sfrac 1 4 \eta^{IL}\theta_{JKL}\epsilon^J_+\epsilon^K_+~,
}
up to terms containing $\A_-$ and  $\epsilon_-$  on the right-hand sides of the above equations; such terms will eventually drop out by setting the corresponding fields to zero, but this has to be done in a consistent way. The quantity $\theta_{IJK}$ is defined as
\ali{
\theta_{IJK}&=-A_{IJK}+3\eta_{L[K}B_{IJ]}{}^L-4\eta_{L[I}B_{J]K}{}^L-3\eta_{L[I}\eta_{\underline{M}J}C_{K]}{}^{LM}-4\eta_{KL}\eta_{M[I}C_{J]}{}^{ML}+\eta_{KL}\eta_{IM}\eta_{JN}D^{MNL}~,
}
with $A$, $B$, $C$ and $D$ being the components of $T_{\hat{I}\hat{J}\hat{K}}$ in \eq{Tmatrix}.

The requirement that the projection onto $L_+$ be well-defined with respect to the BRST symmetry means that the transformations of $\A_-$ and $\epsilon_-$ must vanish. Therefore, setting $\delta \A_-=\delta\epsilon_-=0$ leads to the fixing of the ghost fields $t_I$ and $v_I$:
\begin{alignat}{2}
v_I&=-\sfrac 1 2 \eta_{IL}\eta^{NM}\rho^{L}_{-M}\theta_{JKN}\epsilon^J_+\epsilon_+^K=:\sfrac 1 2 \Theta_{IJK}(\X)\epsilon_+^J\epsilon_+^K\label{vfix}~,\\[4pt]
t_I&=
\Theta_{IJK}(\X)\A_+^J\epsilon_+^K+\sfrac 1 2 \partial_K\Theta_{ILM}F^{\dagger K}\epsilon_+^L\epsilon_+^M~.\label{tfix}
\end{alignat}
We used the fact that $\rho^I_{-J}$ satisfy (3.10), since one can write
\be
0=\eta^{\hat{I}\hat{J}}\rho^K{}_{\hat I}\rho^L{}_{\hat J}=\sfrac 1 2 \eta^{IJ}\left(\rho_{+J}^K\rho_{+I}^L-\rho_{-J}^K\rho_{-I}^L\right)~.
\ee
Fixing of the ghosts $t$ and $v$ is a consequence of choosing the  map $\rho_+$ as in \eqref{dftrho}. Recall that the anchor map of an exact Courant algebroid has a kernel; in the standard case of the projection to the tangent bundle it is all of the cotangent bundle. However, a DFT algebroid is different and this can be seen as follows. Choosing the above parametrization for $\rho_+$, this map has no kernel and therefore we 
have to fix the symmetry associated to the gauge parameter $t$ that came from the Courant algebroid where the map had a kernel instead.  
As we have the fixed ghosts $t_I$ and $v_I$, their BRST transformations must be consistent with those coming from the master action \eq{vart} and \eq{varv}. Applying the BRST operator on \eq{tfix} one obtains:
\ali{
\delta t_I&=\partial_L\Theta_{IJK}\delta\X^L \A^J_+\epsilon_+^K+\Theta_{IJK}\delta \A^J_+\epsilon^K_++\Theta_{IJK} \A^J_+\delta\epsilon^K_++\sfrac 1 2 \partial_A\partial_K\Theta_{ILM}\delta\X^AF^{\dagger K}\epsilon_+^L\epsilon_+^M+{}\nn\\
&\phantom{\,=\,}+\sfrac 1 2 \partial_K\Theta_{ILM}\delta F^{\dagger K}\epsilon_+^L\epsilon_+^M-\partial_K\Theta_{ILM}F^{\dagger K}\delta\epsilon_+^L\epsilon_+^M\nn\\
&=\Theta_{IJK}\dd\epsilon_+^J\epsilon_+^K+\sfrac 1 2 \partial_K\Theta_{ILM}{\cal D}\X^K\epsilon_+^L\epsilon_+^M+\sfrac 1 4 \Big(\eta^{JL}\rho_{+L}^K\Theta_{IJD}\partial_A\Theta_{KBC}+{}\nn\\
&\phantom{\,=\,} +\partial_A\big(2\eta^{JL}\Theta_{IJD}\hat T_{LBC}-2\partial_J\Theta_{ICD}\rho_{+B}^J-\eta^{JL}\rho_{+L}^K\Theta_{IJD}\Theta_{KBC}\big)\Big)F^{\dagger A}\epsilon_+^B\epsilon_+^C\epsilon_+^D+{}\nn\\
&\phantom{\,=\,}+\Big(\partial_M\Theta_{IJL}\rho_{+K}^M-\sfrac 1 2 \Theta_{IPL}\eta^{PR}\rho_{+R}^N\Theta_{NJK}+ \Theta_{IML}\eta^{MN}\hat T_{NJK}-\sfrac 1 2 \Theta_{IMJ}\eta^{MN}\hat T_{NKL}+{}\nn\\
&\phantom{\,=\,+\Big(}+\sfrac 1 4 \Theta_{IJM}\eta^{MN}\rho_{+N}^P\Theta_{PKL}\Big)\A^J_+\epsilon^K_+\epsilon^L_+~\label{deltatprime},
}
using the BRST transformation for $F^{\dagger I}$:
\be
\delta F^{\dagger I}={\cal D}\X^I-\partial_J\rho_{+K}^IF^{\dagger J}\epsilon_+^K~\label{deltafdagger}.
\ee
However, the projection of \eq{vart} implies the following transformation:
\ali{
\delta t_I&=\dd v_I-\epsilon_+^J\partial_I\rho_{+J}^Kt_K+\A^J_+\partial_I\rho_{+J}^Kv_K- \partial_I \hat T_{JKL}\epsilon_+^J\epsilon_+^K\A^L_+-\partial_I\partial_J\rho_{+L}^K\epsilon_+^LF^{\dagger J}v_K-{}\nn\\
&\phantom{\,=\,}-\sfrac 1 3\partial_I\partial_J\hat T_{KLM}F^{\dagger J}\epsilon_+^K\epsilon_+^L\epsilon_+^M\nn\\[4pt]
&=\sfrac 1 2 \partial_J\Theta_{IKL} {\cal D}\X^J\epsilon_+^K\epsilon_+^L+\Theta_{IJK}\dd\epsilon_+^J\epsilon^L_++{}\nn\\
&\phantom{\,=\,}+\Big(\sfrac 1 2 \partial_M\Theta_{IKL}\rho_{+J}^M-\partial_I\rho_{+K}^M\Theta_{MJL}+\sfrac 1 2 \partial_I\rho_{+J}^M\Theta_{MKL}- \partial_I \hat T_{JKL}\Big)\A^J_+\epsilon_+^K\epsilon_+^L~\label{deltats}.
}
Eqs. \eq{deltatprime} and \eq{deltats} should coincide. Therefore, the consistency condition is:
\bea
3 S_{IJKL}\A_+^{J}\epsilon_+^K\epsilon_+^L+ \partial_AS_{IBCD}F^{\dagger A}\epsilon_+^B\epsilon_+^C\epsilon_+^D-\sfrac 1 2 R^K_{\phantom{K}IB}\partial_A\Theta_{KCD}F^{\dagger A}\epsilon_+^B\epsilon_+^C\epsilon_+^D=0\label{condt}~,
\eea
where,
\ali{
S_{IJKL}&:=\partial_M\Theta_{I[JK}\rho_{+L]}^M- \Theta_{IM[J}\eta^{MN}\hat T_{\underline{N}KL]}+\sfrac 1 2 \eta^{MN}\rho_{+N}^P\Theta_{IM[J}\Theta_{\underline{P}KL]}-\sfrac 2 3 \partial_I \hat T_{JKL}+\partial_I\rho_{+[J}^M\Theta_{\underline{M}KL]}~,\\
R^I_{\phantom{I}JK}&:=\eta^{AB}\rho_{+B}^I\Theta_{JAK}+2\partial_J\rho_{+K}^I~.\label{defR}
}
The same can be done for ghost $v_I$ and we obtain:
\bea
S_{IJKL}\epsilon_+^J\epsilon_+^K\epsilon_+^L=0~\label{condv}.
\eea
Fixing function $\Theta_{IJK}(\X)$ by setting:
\be
R^I_{\phantom{I}JK}=0~\label{R=0},
\ee
in \eq{defR} can be shown to imply $S_{IJKL}=0$ meaning conditions \eq{condt} and \eq{condv} are automatically satisfied.
%

\subsection{Projected gauge transformations}

Once we consistently projected all components of the superfields we obtain the following set of gauge transformations:\footnote{From now on we denote $\A_+=A$ and drop all other $\pm$ subscripts.}
\bea\label{gt2}
&&\delta_\epsilon \X^I=\r^I{}_J\epsilon^J~,\\[4pt]
&&\delta_\epsilon A^I=\dd\epsilon^I+\Phi^I{}_{JK}A^J\e^K~, \\[4pt]
&&\d_{\epsilon}F_I=-\dd(\Theta_{IJK}A^J\epsilon^K)-\epsilon^J\partial_I\rho^{K}{}_{J}F_K+\epsilon^JA^K\w A^L(\partial_I \hat T_{KLJ}-\partial_I\r^ N{}_K\Theta_{NLJ})~,
\eea
where we defined
\be 
\Phi^I{}_{JK}:=\eta^{IN}(\hat T_{NJK}-\sfrac 12\r^M{}_N\Theta_{MJK})~.
\ee 
Note that the gauge variation of $F_I$ now includes trivial gauge transformations proportional to the equations of motion. 
 
As we did for the Courant sigma-model case, we examine the transformation of the field equations obtained by varying the action \eqref{dft1}
with respect to $F_{I}, A^{{I}}$ and $\mathbb{X}^{I}$ respectively
\bea\label{DFTeoms}
\label{Feom2} && {\cal D}\X^I:= \dd\X^I-\r^I{}_{J}\, A^{J}=0~,\\[4pt]
\label{Aeom2} && {\cal D}A^{{I}}:=\dd A^{I}-\sfrac 12 \eta^{IK}\rho^J{}_{K}F_J+ \sfrac 1 2  \eta^{IK}\hat T_{ KJL}A^{J}\w A^{L}=0~, \\[4pt]
\label{Xeom2} && {\cal D}F_{I}:=\dd F_I+\partial_I\r^J{}_K \,A^{K}\w F_J-\sfrac 13 \partial_I\hat T_{JKL}\,A^{J}\w A^{K}\w A^{L}=0~.\eea
The gauge transformation of the first field equation gives:
\bea\label{ef1}
\delta_\epsilon {\cal D}\X^I=\epsilon^J\partial_M\r^I{}_J {\cal D}\X^M+\epsilon^J A^K(2\r^M{}_{[K}\partial_{\underline M}\rho^I{}_{J]}-\r^I{}_N\Phi^N{}_{KJ})~.\eea
Therefore, the first condition from the covariance of the field equation is
\be\label{cd2}
2\r^M{}_{[K}\partial_{\underline M}\rho^I{}_{J]}-\r^I{}_N\Phi^N{}_{KJ}=0~.\ee
If we compare this expression with the DFT fluxes obtained by twisting the C-bracket \eqref{cd3a},
 we obtain
 \be\label{gftt}
 \Theta_{NKJ}(X)=-2 \eta_{IN}\r_{M[K}\partial^I\r^M{}_{J]}~,\ee
which is precisely the fixing \eq{R=0}. Next we check the transformation of the field equation of $A^I$ and obtain:
\bea\label{aeom}
\delta_\epsilon{\cal D} A^I&=&\eta^{IN}(\partial_M\hat T_{NJK}-\sfrac 12\partial_M\r^L{}_N\Theta_{LJK}) \epsilon^K{\cal D}\X^M\w A^J+\eta^{IN}\hat T_{NJK}\epsilon^K {\cal D} A^J+{}\nn\\[4pt]
&&+\,\sfrac 12\eta^{IN}\e^KA^M\w A^{M'}\left(3\r^J{}_{[K}\partial_{\underline J}\hat T_{MM']N}-\r^J{}_N\partial_J\hat T_{[KMM']}-3\eta^{PJ}\hat T_{P[MM'}\hat T_{K]NJ}\right)+{}\nn\\[4pt]
&&+\,\sfrac 14\eta^{IN}\underline{\r^P{}_J\eta^{JL}\r^S{}_L}\e^K\left(\Theta_{PKN} F_S+\Theta_{SMN}\Theta_{PM'K}A^M\w A^{M'}\right)
~.\eea
Here the underlined contribution is highlighted for later reference, as it would vanish in the case of a Courant algebroid. We see that the gauge variation of the field equation of $A$ is covariant provided that
\be\label{cd3}
3\r^J{}_{[K}\partial_{\underline J}\hat T_{MM']N}-\r^J{}_N\partial_J\hat T_{KMM'}-3\eta^{PJ}\hat T_{P[MM'}\hat T_{K]NJ}=0~.\ee
This is one of the local coordinate expressions for a DFT algebroid. 
However, due to \eqref{cd1}, the last line in \eqref{aeom} does not vanish, thus there is an additional obstruction. Let us look at this obstruction in more detail: 
 \bea\label{odft1}
 &&\eta^{IN}\underline{\r^P{}_J\eta^{JL}\r^S{}_L}\e^K\left(\Theta_{PKN} F_S+\Theta_{SMN}\Theta_{PM'K}A^M\w A^{M'}\right)={}\nn\\
 &&{}={}\eta^{IN}\epsilon^K\left(\eta^{PS}\Theta_{PKN} F_S+\Theta_{SMN}\eta^{PS}\Theta_{PM'K}A^M\w A^{M'}\right)~.\eea
 The first term in the parentheses can be rewritten using \eqref{gftt} as
 \be
 \eta^{PS}\Theta_{PKN} F_S=-2\r_{M[K}\partial^S\r^M{}_{N]}F_S~, \ee
 which vanishes due to the already imposed condition \eqref{anomaly}.  The second term in the round brackets gives explicitly
 \be\label{exst}
 \Theta_{SMN}\eta^{PS}\Theta_{PM'K}=4\eta_{SS'}\r_{L[M}\partial^{S'}\r^L{}_{N]}\r_{J[M'}\partial^S\r^J{}_{K]}~,\ee
 again after using \eqref{gftt}. This term has precisely the form of the DFT strong constraint. 

What about closure of the algebra of gauge transformations? On $\X^I$ we have
\bea\label{cX}
[\delta_{\epsilon_1},\delta_{\epsilon_2}]\X^I&=&\r^I{}_J\epsilon_{12}^J~,\\[4pt] \epsilon_{12}^I&:=&\Phi^I{}_{KL}\epsilon_1^K\epsilon_2^L~,\eea
where we used the  condition  \eqref{cd2} to define $\epsilon_{12}$.   On $A^I$  we have:
\ali{\label{cA}
[\delta_{\epsilon_1},\delta_{\epsilon_2}]A^I&=\delta_{\e_{12}}A^I-\partial_L\Phi^I{}_{JK}\e_1^J\e_2^K {\cal D}\X^L+{}\nn\\[4pt]
&\phantom{\,=\,}+3\left(\Phi^I{}_{N[M}\Phi^N{}_{JK]}-\r^N{}_{[M}\partial_{\underline{N}}\Phi^I{}_{JK]}\right)\e_1^J\e_2^K A^M
~,}
where we used \eqref{cd2} and \eqref{cd3}. The last   line vanishes identically  using  \eqref{cd2}, thus we have  the on-shell  closure of the  algebra of gauge transformations. However, we obtain consistent  gauge transformations of the field equations   only after applying the  strong constraint, c.f. the underlined term in Eq. \eqref{aeom}.

\section{Conclusion and outlook}

We have shown how  to construct the gauge symmetry of the DFT worldvolume action by projecting the superfield components and BRST transformations of a Courant sigma-model master BV action defined over doubled space. We obtained that the algebra of gauge transformations  closes on-shell.  However, the field equations transform covariantly only upon the use of a constraint, which is the analogue of the DFT strong constraint. This is in accord with the statement that  the target space DFT  action is invariant under the generalized diffeomorphisms only after using the strong constraint. Our approach establishes this result at the level of the worldvolume theory.

An interesting  question which remains open is whether one  can  find a modification of the DFT  worldvolume action \eqref{dft1} and/or \eqref{masterDFT}  in order to achieve  gauge invariance without the use of the strong constraint. There are two main reasons why one should  attempt to construct such an  improved DFT action. The first reason is that conjectured non-commutative and non-associative closed string backgrounds do not satisfy the strong constraint, and therefore one presumably needs to go beyond DFT in order to consistently describe such backgrounds.   The other reason is that one would like to use the AKSZ construction in order to obtain an action satisfying the classical master equation. This would be a first step toward quantization of the (improved) DFT sigma-model.

\acknowledgments
We acknowledge support by COST (European Cooperation in Science and Technology) in the framework of the Action MP1405 QSPACE.
The work of C. G, L.J. and F.S.K. was supported by the Croatian Science Foundation under the Project IP-2014-09-3258.
A.Ch., C. G., L.J. and F.S.K. were partially supported by
the European Union through the European Regional Development Fund - the Competitiveness and Cohesion Operational Programme (KK.01.1.1.06).
The work of R.J.S. was supported by the Consolidated Grant ST/P000363/1 from the U.K. Science and Technology Facilities Council.

\end{document}